\documentclass[runningheads]{llncs}
\usepackage{url}

\usepackage{graphicx}
\usepackage{caption}
\usepackage{multirow}
\usepackage{float}
\usepackage{mathptmx}
\usepackage{anyfontsize}
\usepackage{t1enc}
\usepackage{subcaption}
\usepackage{threeparttablex}
\usepackage{booktabs}
\usepackage{array}
\usepackage{amssymb}
\usepackage{bbding}
\begin{document}

\title{HPC AI500: A Benchmark Suite for HPC AI Systems}
%
%
\author{Zihan Jiang\inst{1,2} \and
Wanling Gao\inst{1,2,3} \and
Lei Wang\inst{1,3} \and
Xingwang Xiong\inst{1,2} \and
Yuchen Zhang\inst{5} \and
Xu Wen\inst{1,2} \and
Chunjie Luo\inst{1} \and
Hainan Ye\inst{3,4} \and
Xiaoyi Lu\inst{6} \and
Yunquan Zhang\inst{9} \and
Shengzhong Feng\inst{7} \and
Kenli Li\inst{8} \and
Weijia Xu\inst{10} \and
Jianfeng Zhan\inst{1,2,3}
\thanks{Jianfeng Zhan is the corresponding author.}
}
\authorrunning{ZH. Jiang et al.}
%
\institute{State Key Laboratory of Computer Architecture, Institute of Computing Technology, Chinese Academy of Sciences \and
University of Chinese Academy of Sciences, China \and
BenchCouncil (International Open Benchmarking Council) \and
Beijing Academy of Frontier Sciences and Technology \and
State University of New York at Buffalo  \and
Department of Computer Science and Engineering, The Ohio State University \and
National Supercomputing Center in Shenzhen, China \and
National Supercomputing Center in Changsha, China \and
National Supercomputing Center in Jinan, China \and
Texas Advanced Computing Center, The Texas University at Austin 
\email{\{jiangzihan,gaowanling, zhanjianfeng,wanglei\_2011,xiongxingwang,wenxu,luochunjie,\\gaowanling,zyq\}@ict.ac.cn}\\
\email{\{zhang232\}@buffalo.edu}\\
\email{\{fengsz\}@nsccsz.cn}\\
\email{\{lkl\}@hnu.edu.cn}\\
\email{\{xwj\}@tacc.utexas.edu}\\
\email{\{yehainan\}@mail.bafst.com}\\
\email{\{luxi\}@cse.ohio-state.edu}}
%
\maketitle              
\begin{abstract}
In recent years, 
with the trend of applying deep learning (DL) in high performance scientific computing,
the unique characteristics of emerging DL workloads in HPC raise great challenges in designing, implementing HPC AI systems. The community needs a new yard stick for evaluating the future HPC systems.
In this paper, we propose HPC AI500 --- a benchmark suite for evaluating HPC systems that running
scientific DL workloads. Covering the most representative scientific fields, each workload from HPC AI500 is based on real-world scientific DL applications. Currently, we choose 14 scientific DL benchmarks from perspectives of application scenarios, data sets, and software stack. We propose a set of metrics for comprehensively evaluating the HPC AI systems, considering both accuracy, performance as well as power and cost. 
 We provide a scalable reference implementation of HPC AI500. The specification and source code are publicly available from \url{http://www.benchcouncil.org/HPCAI500/index.html}. Meanwhile, the AI benchmark suites for datacenter, IoT, Edge are also released on the BenchCouncil web site.

\end{abstract}
~\\
\section{Introduction}~\label{Introduction}

The huge success of AlexNet~\cite{alexnet} in the ImageNet~\cite{imagenetweb} competition marks that deep learning(DL) is leading the renaissance of Artificial Intelligence (AI). Since then, a wide range of application areas have started using DL and achieved unprecedented results, such as image recognition, natural language processing, and even autonomous driving. In the commercial fields, many DL-based novel applications have emerged, creating huge economic benefits. In the fields of high performance scientific computing, similar classes of problems are faced, i.e., predicting extreme weather~\cite{weather}, finding signals of new particles~\cite{high-energy}, and estimating cosmological parameters~\cite{cosmology}. These scientific fields are essentially solving the common class of problems that exist in commercial fields such as classifying images, predicting classes labels, or regressing a numerical quantity. In several scientific computing fields, DL has replaced traditional scientific computing methods and becomes a promising tool~\cite{deeplearninginscience}.

	As an emerging workload in high performance scientific computing, DL has many unique features compared to traditional high performance computing. First, training a DL model depends on massive data that are represented by high-dimensional matrices. Second, leveraging deep learning frameworks such as Tensorflow~\cite{tensorflow} and caffe~\cite{caffe} aggravates the difficulty of the software and hardware co-design. Last but not least, the heterogeneous computing platform of DL is far more complicated than traditional scientific workloads, including CPU, GPU, and various domain-specific processor (e.g. Cambricon Diannao~\cite{diannao} or Google TPU \cite{tpu}). Consequently, the community requires a new yardstick for evaluating future HPC AI systems. However, the diversity of scientific DL workloads raise great challenges in HPC AI benchmarking.

	\begin{enumerate}
		\item Dataset: Scientific data is often more complex than MINST or ImageNet data sets. First, the shape of scientific data can be 2D images or higher-dimension structures. Second, there are hundreds of channels in a scientific image, while the popular image data often consists of only RGB. Third, Scientific datasets are always terabytes or even petabytes in size.
		\item Workloads: Modern scientific DL doesn't adopt off-the-shelf models, instead builds more complex model with domain scientific principles (e.g. energy conservation)~\cite{weather}.
		\item Metrics: Due to the importance of accuracy, using a single performance metric such as FLOPS leads to insufficient evaluation. For a comprehensively evaluation, the selected metrics should not only consider the performance of the system, but also consider the accuracy of the DL model~\cite{dawnbench}.
		\item Scalability: Since the scientific DL workloads always run on the supercomputers, which are equipped with tens of thousands nodes, the benchmark program must be highly scalable.
	\end{enumerate}
	
Most of the existing AI benchmarks~\cite{dawnbench,MLPerf,TBD,fathom,BenchNN,ccbdbench} are based on commercial scenarios. Deep500~\cite{deep500} is a benchmarking framework aiming to evaluate high-performance deep learning. However, its reference implementation uses commercial open source data sets and simple DL models, hence cannot reflect real-world HPC AI workloads. We summary these major benchmarking efforts for AI and compare them with HPC AI500 as shown in the table below.

\begin{table}
\begin{threeparttable}

\caption {Comparison of AI Benchmarking Efforts.}
\begin{tabular}{|c|c|c|c|c|c|c|c|}
\hline
\multirow{3}{*}{Benchmark Efforts} &
\multirow{3}{*}{Datasets} &
\multicolumn{4}{c|}{Problem domains} &
\multicolumn{2}{c|}{Implementation} \\
\cline{3-8}
 & & \multicolumn{3}{c|}{Scientific} & \multirow{2}{*}{Commercial}& \multirow{2}{*}{Standalone} & \multirow{2}{*}{Distributed} \\
 
\cline{3-5}
& & EWA\tnote{1}\,\, & Cos\tnote{2}\,\, & HEP\tnote{3}\,\, & & & \\

\cline{1-8}
HPC AI500 & Scientific data & $\checkmark$ & $\checkmark$ & $\checkmark$ & $\times$ & $\checkmark$ & $\checkmark$  \\

\hline
TBD & Commercial data & $\times$ & $\times$ & $\times$ & $\checkmark$ & $\checkmark$ & $\times$ \\
\hline
MLPerf & Commercial data & $\times$ & $\times$ & $\times$ & $\checkmark$ & $\checkmark$ & $\times$ \\
\hline
DAWNBench & Commercial data & $\times$ & $\times$ & $\times$ & $\checkmark$ & $\checkmark$ & $\times$ \\
\hline
Fathom & Commercial data & $\times$ & $\times$ & $\times$ & $\checkmark$ & $\checkmark$ & $\times$ \\
\hline
Deep500 & Commercial data & \multicolumn{4}{c|}{Framework, undefined} &  $\checkmark$ & $\checkmark$ \\
\hline
\end{tabular}
\begin{tablenotes}
\footnotesize
        \item[1] Extreme Weather Analysis \item[2] Cosmology \item[3] High Energy Physics 

\end{tablenotes}

\end{threeparttable}
\end{table}
	Consequently, targeting above challenges, we propose HPC AI500---a benchmark suite for HPC AI systems. Our major contributions are as follows:
		\begin{enumerate}
		\item We create a new benchmark suite that covers the major areas of high performance scientific computing. The benchmark suite consists of micro benchmarks  and component benchmarks. The workloads from component benchmarks use the state-of-the-art models and representative scientific data sets to reflect the real-world performance results. In addition, we select several DL kernels as the micro benchmarks for evaluating the upper bound performance of the systems.
	    \item We propose a set of metrics for comprehensively evaluating the HPC AI systems. 
 Our metrics for component benchmarks include both accuracy and performance. For micro benchmarks, we provide metrics such as FLOPS to reflect the upper bound performance of the system.
	\end{enumerate}

Coordinated by BenchCouncil (\url{http://www.benchcouncil.org}), we also release the datacenter AI benchmarks~\cite{datacenter,aibench}, the IoT AI benchmarks~\cite{IoT}, edge AI benchmarks~\cite{edge}, and big data benchmarks~\cite{bdb,dcbench}, which are publicly available from \url{http://www.benchcouncil.org/HPCAI500/index.html}.

\section{Deep Learning in Scientific Computing} \label{ScientificDL}
In order to benchmark HPC AI systems, the first step is to figure out how DL works in scientific fields. Although it is an emerging field, several scientific fields have applied DL to solve many important problems, such as extreme weather analysis ~\cite{climatedataset,Globenet,extremeweather,weather}, high energy physics ~\cite{trackfinding,cellularHEP,jet-images,jetdiscrimination,high-energy}, and cosmology ~\cite{cosmology,cosmology2,galaxyimage,universerGAN,cosmologicalmodel}.

\subsection {Extreme Weather Analysis}
Extreme Weather Analysis (EWA) poses a great challenge to human society. It brings severe damage to people health and economy every single year. For instance, the heatwaves in 2018 caused over 1600 deaths according to the UN report~\cite{heatwave}. And the landfall of hurricane Florence and Michael caused about 40 billion dollars worth of damage to US economy~\cite{hurricane}. In this context, understanding extreme weather life cycle and even predicting its future trend become a significant scientific goal. Achieving this goal always requires accurately identifying the weather patterns to acquire the insight of climate change based on massive climate data analysis. Traditional climate data analysis methods are built upon human expertise in defining multi-variate thresholds of extreme weather events. However, it has a major drawback: there is no commonly held set of criteria that can define a weather event due to the man-made subjectivism, which leads to inaccurate pattern extraction. Therefore, DL has become another option for climate scientists. Liu et al. (2016)~\cite{climatedataset} develop a relatively simple CNN model with two convolutional layers to classify three typical extreme weather events and achieve up to 99\% accuracy. Racah et al. (2017)~\cite{extremeweather} implement a multichannel spatiotemporal CNN architecture for semi-supervised  prediction and exploratory extreme weather data analysis. GlobeNet~\cite{Globenet} is a CNN model with inception units for typhoon eye tracking. Kurth et al. (2018)~\cite{weather} use variants of Tiramisu and DeepLabv3+ neural networks which are both built on Residual Network (ResNet)~\cite{resnet}. They deployed these two networks on Summit and firstly achieved exascale deep learning for climate analysis.

 \subsection {High Energy Physics}
	Particle collision is the most important experiment approach in High Energy Physics (HEP). Detecting the signal of new particle is the major goal in experimental HEP. Today's HEP experimental facility such as LHC creates particle signals with hundreds of millions channels with a high data rate. The signal data from different channels in every collision usually  are represented as a sparse 2d image, so called a jet-image. In fact, accurately classifying these jet-images is the key to find signals of new particles.
	In recent years, due to the excellent performance in pattern recognition, DL has become the focus of the data scientists in HEP community and has a tendency to go mainstream. Oliveira et al. (2016)~\cite{jet-images} use a CNN model with 3 convolutional layers to tag jet-images. They firstly demonstrated that using DL not only improve the discrimination power, but also gain new insights compared to designing physics-inspired features. Komiske et al. (2017)~\cite{jetdiscrimination} adopt a CNN model to discriminate quark and gluon jet-image. Kurth et al.(2017)~\cite{high-energy} successfully deploy CNN to analyze massive HEP data on the HPC system and achieve petaflops performance. Their work is the first attempt at scaling DL on large-scale HPC systems.

\subsection {Cosmology}
Cosmology is a branch of astronomy concerned with the studies of the origin and evolution of the universe, from the Big Bang to today and on into the future~\cite{cosmologydef}. In 21st century, the most fundamental problem in cosmology is the nature of dark energy. However, this mysterious energy greatly affects the distribution of matter in the universe that is described by cosmological parameters. Thus, accurately estimating these parameters is the key to understand the insight of the dark energy. For solving this problem, Ravanbakhsh et al. (2017)~\cite{cosmology2} firstly propose a 3D CNN model with 6 convolutional layers and 3 fully-connected layers and opens the way to estimating the parameters with high accuracy. Mathuriya et al. (2018) propose CosmoFlow~\cite{cosmology}, which is a project aiming to process large 3D cosmology dataset on HPC systems. They extend the CNN model designed by  Ravanbakhsh et al. (2017)~\cite{cosmology2}. Meanwhile, in order to guarantee the high fidelity numerical simulations and avoid the use of expensive instruments, generating high quality cosmological data is also important.
 Ravanbakhsh et al. (2017)~\cite{galaxyimage} propose a deep generative model for acquiring high quality galaxy images. Their results show a reliable alternative for generating the calibration data of cosmological surveys.

\subsection {Summary}
After investigating the above representative scientific fields, we have identified the representative DL applications and abstracted these DL applications into classical AI tasks. As shown in Table~\ref{table:scientificDL}, almost all the applications are essentially using CNN to extract the patterns of various scientific image data.  From this perspective, \textit{image recognition}, \textit{image generation}, and \textit{object detection} are the most important tasks in modern scientific DL. In our benchmark methodology  (Section~\ref{meth}), we use these three classic AI tasks as the component workloads of the HPC AI500 Benchmark.

\begin{table}
\caption{Modern Scientific Deep Learning.}	
\begin{threeparttable}

\begin{tabular}{|c|c|c|c|}
\hline
\centering
\textbf{Scientific Fields} & \textbf{DL Applications} & \textbf{Classical DL Tasks} & \textbf{Model Type} \\
\hline
\centering
Extreme Weather Analysis & Identify weather patterns & Object Detection & CNN\\[1.2ex]
\hline
\centering
High Energy Physics & Jet-images discrimination & Image Recognition & CNN\\[1.2ex]
\hline
\centering
Cosmology
&

\begin{tabular}{@{}c@{}}
 Estimate parameters\\

 Galaxy image generation\\	
\end{tabular}
&
\begin{tabular}{@{}c@{}}
 Image Recognition\\

 Image Generation\\	
\end{tabular}
& CNN
\\
\hline

\end{tabular}
	
\end{threeparttable}
\label{table:scientificDL}
\end{table}

\section{Benchmarking Methodology and Decisions} \label{Benchmark}

\subsection{Methodology} \label{meth}
\begin{figure}

\includegraphics[width=\textwidth]{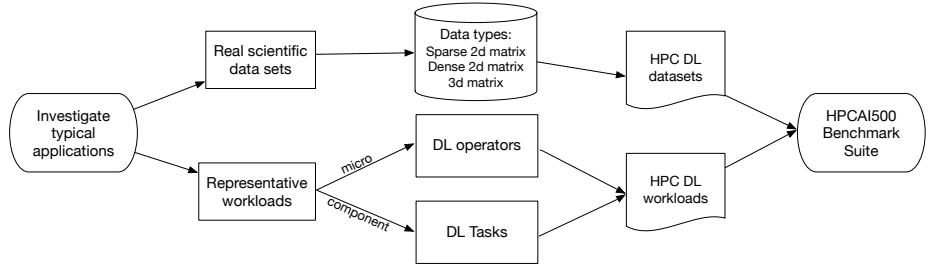}
\centering
\caption{HPCAI500 Methodology}
\label{methodology}
\end{figure}

Our benchmarking methodology is shown in Figure \ref{methodology}, similar to that~\cite{bdb}. As HPC AI is an emerging and evolving domain, we take an incremental and iterative approach. First of all, we investigate the scientific fields that use DL widely. As mentioned in Section~\ref{ScientificDL}, \textit{extreme weather analysis, high energy physics}, and \textit{cosmology} are the most representative fields. Then, we pay attention to the typical DL workloads and data sets in these three application fields.

In order to cover the diversity of workloads,
we focus on the critical tasks that DL has performed in the aforementioned fields. Based on our analysis in Section~\ref{ScientificDL}, we extracts three important component benchmarks that can represent modern scientific DL, namely \textit{image recognition}, \textit{image generation}, and \textit{object detection}. This shows that CNN models play an important role. In each component, we choose the state-of-the-art model and software stack from the applications. We also select the hotspot DL operators as the micro benchmark for evaluating upper bound performance of the system.

We chose three real-world scientific data sets from aforementioned scientific fields and consider their diversity from the perspective of data formats. In modern DL, the raw data is always transformed into matrix for downstream processing. Therefore, we classify these matrices into three kinds of formats: 2D sparse matrix, 2D dense matrix, and 3 dimensional matrix. In each matrix format, we also consider the unique characteristics (e.g., multichannel that more than RGB, high resolution) in the scientific data.

\subsection{The Selected Datasets}

We investigate the representative data sets in our selected scientific fields and collect three data sets as
shown in Table~\ref{table:datasets}. Our selection guidelines follow the aforementioned benchmarking methodology.

\textbf{The Extreme Weather Data set}~\cite{weatherdata} is made up of 26-year of climate data. The data of every year is available as one HDF5 file. Each HDF5 file contains two data sets: images and boxes. \emph{Images data set has 1460 example dense images} (4 per day, 365 days per year) with 16 channels. Each channel is 768 * 1152 corresponding to one measurement per 25 square km on earth. Boxes dataset records the coordinates of the four extreme weather events in the corresponding images: tropical depression, tropical cyclone, extratropical cyclone and the atmospheric river.

\textbf{The HEP Data set}~\cite{smalljetimages} is divided into two classes: the RPV-Susy signal and the most prevalent background. The training data set is composed of around 400 k jet-images. Each jet-image is represented as a 64*64 sparse matrix and has 3 channels. It also provides validation and test data. All the data are generated by using the Pythia event generator~\cite{pythia} interfaced to \emph{the Delphes fast detector simulation}~\cite{jet-images}.

\textbf{The Cosmology Data set}~\cite{cosmology} aims to predict the parameters of cosmology. It is based on dark matter N-body simulations produced using the MUSIC~\cite{MUSIC} and pycola~\cite{pycola} packages. Each simulation covers the volumes of \(512h^{-1}Mpc^3\) and contains \(512^3\) dark matter particles.

\begin{table}[h]
	\caption{The Chosen Datasets}
	\centering
	
	\begin{tabular}{|c|c|c|}
	\hline	
		\textbf{Dataset} & \textbf{Data Format} & \textbf{Scientific Features} \\
		\hline
		Extreme Weather Dataset& 2D dense matrix & high resolution, multichannel\\
		
		HEP Dataeset& 2D sparse matrix & multichannel\\
		
		Cosmology Dataset  &  3D matrix & multidimensional\\	
	\hline	
	\end{tabular}
	\label{table:datasets}
\end{table}
\subsection{The Selected Workloads}
\subsubsection{Component Benchmarks}
Since object detection, image recognition, and image generation are the most representative DL tasks in modern scientific DL. We choose the following state-of-the-art models as the HPC AI500 component benchmarks.
\paragraph{Faster-RCNN}~\cite{fasterRCNN} targets real-time object detection. Unlike the previous object detection model ~\cite{RCNN,fastRCNN}, it replaces the selective search by a region proposal network that achieves nearly cost-free region proposals. Further more, Faster-RCNN combines the advanced CNN model as their base network for extracting features and is the foundation of the 1st-place winning entries in ILSVRC'15 (ImageNet Large Scale Visual Recognition Competition).
\paragraph{ResNet}~\cite{ResNet} is a milestone in Image Recognition, marking the ability of AI to identify images beyond humans. It solves the degradation problem, which means in the very deep neural network the gradient will gradually disappear in the process of propagation, leading to poor performance. Due to the idea of ResNet, researchers successfully build a 152-layer deep CNN. This ultra deep model won all the awards in ILSVRC'15.
\paragraph{DCGAN}~\cite{DCGAN} is one of the popular and successful neural network for GAN~\cite{GAN}. Its fundamental idea is replacing fully connected layers with convolutions and using transposed convolution for upsampling. The proposal of DCGAN helps bride the gap between CNNs for supervised learning and unsupervised learning.

\subsubsection{Micro Benchmarks}
We choose the following primary operators in CNN as our micro benchmarks.
\paragraph{Convolution}
In mathematics, convolution is a mathematical operation on two functions to produce a third function that expresses how the shape of one is modified by the other~\cite{wiki_conv}. In a CNN, convolution is the operation occupying the largest proportion, which is the multiply accumulate of the input matrix and the convolution kernel, and then produces feature maps. There are many convolution kernels distributed in different layers responsible for learning different level features.
\paragraph{Full-connected}
The full-connected layer can be seen as the classifier of a CNN, which is essentially matrix multiplication. It is also the cause of the explosion of CNN parameters. For example, in AlexNet~\cite{alexnet}, the number of training parameters of fully-connected layers reaches about 59 million and accounts for 94 percent of the total. 	
\paragraph{Pooling}
Pooling is a sample-based discretization process. In a CNN, the objective of pooling is to down-sample the inputs (e.g., feature maps), which leads to the reduction of dimensionality and training parameters. In addition, it enhances the robustness of the whole network. The commonly used pooling operations including max-pooling and average-pooling.

\begin{table}
\caption {The Summary of HPC AI500 Benchmark.}
\begin{threeparttable}
\begin{tabular}{c|c|c|c|c|c}
\hline
 \textbf{App Scenarios} & \textbf{Workloads} & \textbf{Fields} & \textbf{Datasets} & \textbf{Data Format} & \textbf{Software Stack}\\
\hline
 Micro Benchmarks  &
 \begin{tabular}{@{}c@{}c@{}}
 Convolution\\
 Pooling\\
 Fully-Connected\\	
 \end{tabular}
 &
 \begin{tabular}{@{}c@{}c@{}}
 HEP\tnote{1}\\
 EWA\tnote{2}\\
 Cos\tnote{3}\\

 \end{tabular} & Matrix &
 \begin{tabular}{@{}c@{}c@{}}
 Sparse 2D Matrix\\
 Dense 2D Matrix\\
 3D Matrix\\	
 \end{tabular}
   &
 \begin{tabular}{@{}c@{}c@{}}
 CUDA\\
 MKL\\	
 \end{tabular}\\

\hline
Image Recognition & ResNet &
\begin{tabular}{@{}c@{}}
 HEP\\

 Cos\\	
 \end{tabular}
&
\begin{tabular}{@{}c@{}}
 HEP Dataset\\
 Cos Dataset\\	
 \end{tabular}
& \begin{tabular}{@{}c@{}}
 Sparse 2D matrix\\
 3D matrix\\	
 \end{tabular}

& \begin{tabular}{@{}c@{}c@{}}
 TensorFlow\\
 Pytorch\\

 \end{tabular}\\

\hline
Object Detection& Faster-RCNN & EWA & EWA Dataset & Dense 2D Matrix &  \begin{tabular}{@{}c@{}c@{}}
 TensorFlow\\
 Pytorch\\

 \end{tabular} \\
\hline
Image Generation & DCGAN & Cos
  &  Cos Dataset& 3D Matrix &
\begin{tabular}{@{}c@{}c@{}}
 TensorFlow\\
 Pytorch\\

 \end{tabular} \\
\hline
\end{tabular}
\begin{tablenotes}
        \footnotesize
        \item[1] High Energy Physics\item[2] Extreme Weather Analysis\item[3] Cosmology
      \end{tablenotes}
\end{threeparttable}
\label{table:summary}
\end{table}

\subsection{Metrics}
\subsubsection{Metrics for Component Benchmarks} At present, time-to-accuracy is the most well-received solution~\cite{dawnbench,MLPerf}. For comprehensive evaluate, the training accuracy and validation accuracy are both provided. The former is used to measure the training effect of the model, and the latter is used to measure the generalization ability of the model. The threshold of target accuracy is defined as a value according to the requirement of corresponding application domains. Each application domain needs to define its own target accuracy. In addition, cost-to-accuracy and power-to-accuracy are provided to measure the money and power spending of training the model to the target accuracy. 

\subsubsection{Metrics for Micro Benchmarks}
The metrics of the micro benchmarks is simple since we only measure the performance without considering accuracy. we adopt FLOPS and images per second(images/s) as two main metrics. We also consider power and cost related metrics.
\section{Reference Implementation}

\subsection{Component Benchmarks}

According to the survey~\cite{survey} of NERSC (National Energy Research Scientific Computing Center, the most representative DL framework is TensorFlow, and the proportion of which is increasing year by year. Consequently, we adopt TensorFlow for preferred framework.
		
		In order to evaluate large-scale HPC systems running scientific DL, scalability is the fundamental requirement. In modern distributed DL, synchronized training through data parallelism is the mainstream. In this training scheme, each training process gets a different portion of the full dataset but has a complete copy of the neural network model. At the end of each batch computation, all processes will synchronize the model parameters by \textit{all\_reduce} operation to ensure they are training a consistent model. TensorFlow implements \textit{all\_reduce} through a parameter server~\cite{ps} and use the GRPC protocol for communication by default. The master-slave architecture and socket-based communication can not extend to large-scale clusters~\cite{tfnotscale}. Horovod~\cite{horovod} iirrespective a library originally designed for scalable distributed deep learning using TensorFlow. It implements \textit{all\_reduce} operation using ring-based algorithm~\cite{ring-based} and MPI (Message Passing Interface) for communication. Due to the decentralized design and high effective protocol, the combination of TensorFlow and Horovod has successfully scaled to 27360 GPUs on Summit~\cite{weather}. Therefore, we leverage Horovod to improve the scalability.

\subsection{Micro Benchmarks}
The goal of micro benchmarks is to determine the upper bound performance of the system. To do so, we implement it with succinct software stack. Every DL operator is written in C++ or call the low-level neural networks library (e.g. CuDNN) without any other dependencies.

\section{Conclusion}
In this paper, we propose HPC AI500---a  benchmark suite for evaluating HPC system that running scientific deep learning workloads. Our benchmarks model real-world scientific deep learning applications, including extreme weather analysis, high energy physics, and cosmology. We propose a set of metrics for comprehensively evaluating the HPC AI systems, considering both accuracy, performance as well as power and cost.
 We provide a scalable reference implementation of HPC AI500. The specification and source code of HPC AI500 are publicly available from \url{http://www.benchcouncil.org/HPCAI500/index.html}.

\section*{Acknowledgment}
This work is supported by the Standardization Research Project of Chinese Academy of Sciences No.BZ201800001.

%
%
%
%

\end{document}